\documentclass[prb,twocolumn,aps]{revtex4}
\usepackage{graphicx}
\usepackage{dcolumn}
\usepackage{bm}
\usepackage{hyperref}

\newcommand{\beq}{\begin{eqnarray}}
\newcommand{\eeq}{\end{eqnarray}}

\begin{document}\title{Kinks and Mid-Infrared Optical Conductivity
from Strong Electron Correlation}
\author{Shiladitya Chakraborty }
\author{Dimitrios Galanakis}
\author{Philip Phillips}
\affiliation{Department of Physics,
University of Illinois
1110 W. Green Street, Urbana, IL 61801, U.S.A.}
\date{\today}
 
\begin{abstract}
   We compute the one-particle spectral function and the optical            
   conductivity for the 2-d Hubbard model on a square lattice. The computational method is cellular dynamical mean-field theory (CDMFT) in which a 4-site Hubbard plaquette is embedded in a self-consistent bath. We
   obtain a `kink' feature in the  dispersion of the spectral function
   and a mid-infrared (mid-IR) absorption peak in the optical
   conductivity, consistent with experimental data. Of the 256
   plaquette states, only a single state which has d$_{x^2-y^2}$
   symmetry  contributes to the
   mid-IR, thereby suggesting a direct link with the pseudogap.  Local
   correlations between doubly and singly occupied sites which lower
   the kinetic energy of a hole are the
   efficient cause of this effect. 
\end {abstract}

\maketitle

A distinctive feature seen in angle-resolved photoemission spectroscopy (ARPES) spectra on the high-temperature copper-oxide superconductors (cuprates) is the `kink' \cite{lanzara} or abrupt change in the slope of the electron dispersion $\omega$ vs $k$.   The abrupt change in the slope in the electron dispersion or the quasiparticle velocity obtained from ARPES, has been observed in various families of hole-doped cuprates  at an energy of 50 meV - 80 meV.  The size of the kink decreases with hole doping x, and the kink seems to disappear around $x = 0.3$.  Since ARPES offers a direct measure of the spectral function, the imaginary part of the self-energy can also be deduced. Experimentally, the imaginary part of the self energy $Im\Sigma(k,\omega)$ shows a suppression, indicating a diminished scattering rate\cite{zhou}, at the kink energy. The presence of the kink is reminiscent of similar observations in the spectra of conventional BCS superconductors, where phonons cause an abrupt change in the electron dispersion as well.
Consequently, explanations based on collective phonon excitations\cite{lanzara,cuk,damascelli,bogdanov,johnson,kaminski,kim} have been invoked to explain the ubiquitous kink-feature. Rival explanations include purely electronic scenarios based on spin-fluctuations\cite{valla,vollhardt}.  Recent ARPES experiments\cite{graf, meevasana} find that kink phenomena occur at high energies as well.  At the high-energy kink (referred to colloquially as the waterfall), the electron dispersion bifurcates.  

Equally striking is the optical conductivity in the mid-infrared. 
The optical conductivity of a material can be  extracted from its
measured reflectivity $R(\omega)$ spectrum using the Kramers- Kronig
(KK) relations\cite{basov1,basov2,opt0}.  Two universal features
seen in the optical conductivity of  the cuprates are: \textbf{1)}a
Drude peak at low energies followed by \textbf{2)}an  absorption
feature in the mid-IR  region. The Drude peak corresponds to the
motion of free carriers in the system for which the Drude model
predicts a conductivity peak at zero frequency.  The existence of an
absorption feature in the mid-IR region is unexpected, since  doped
Mott insulators are expected to have spectral weight either at the
high energy sector across the Mott gap(UV) or at low energy, close to
the Fermi level(far IR), \cite{meinders}. The spectral weight of the
mid-IR band in the optical conductivity is an
increasing\cite{tanner,cooper,uchida,opt1,opt2,opt3}
function of doping whereas the band at high energy (UV) decreases.
This tradeoff suggests that the origin of the mid-IR band arises
fundamentally from the strong correlations that mediate the Mott
state. It is not surprising then that no low $T_c$ materials exhibit a
mid-IR band (MIB) in their optical conductivity. Amidst the myriad explanations proposed\cite{dagotto,lorenzana1,lorenzana2,emery1,emery2,leggett,turlakov},
 none has emerged as the definitive answer to this problem. 

In a series of recent papers\cite{ftm1,ftm2,ftm3} we formulated the
exact low-energy theory of a doped Mott insulator and showed that a
new excitation emerges at low energies.  The excitation is a charge 2e
boson which is not made out of elemental excitations.  This excitation
was shown to offer a unifying mechanism for the kink features (both
high and low energy as well as the bifurcation of the electron
dispersion) and the mid-IR band.  What we present here are a series of
numerical studies on the Hubbard model which offers an independent
test of the predictions of the low-energy theory.  To this end, we
employ the state-of-the-art numerical method on the Hubbard model,
cellular dynamical mean-field theory (CDMFT)\cite{kotliar}, to investigate whether
the low-energy kink and mid-IR bands have purely electronic origins.
 We find that both the
kink and the mid-IR are linked to short-range electronic
correlations.  In particular, the mid-IR arises from a mixing of the
upper and lower Hubbard bands through an electronic state that has
d$_{x^2-y^2}$ symmetry, suggesting that the pseudogap is also due to the same mechanism.   This mechanism is identical to that mediated
by the charge 2e boson in the exact low-energy theory.  

To this end, we start with the 2D Hubbard model,
\beq
H_{\rm Hubb}&=&-t\sum_{i,j,\sigma} g_{ij} c^\dagger_{i,\sigma}c_{j,\sigma}+U\sum_{i,\sigma} c^\dagger_{i,\uparrow}c^\dagger_{i,\downarrow}c_{i,\downarrow}c_{i,\uparrow}
\eeq
. Here $i,j$ label lattice sites, $g_{ij}$ is equal to one if $i,j$
are nearest neighbours, $c_{i\sigma}$ annihilates an electron with
spin $\sigma$ on lattice site $i$, $t$ is the nearest-neighbour
hopping matrix element and $U$ the energy cost when two electrons
doubly occupy the same site. The cuprates live in the strongly coupled
regime in which the interactions dominate as $t\approx 0.5$eV and
$U=4$eV. Various  numerical techniques have been developed to study
strongly interacting systems on a lattice. These include exact
diagonalization (Lanczos technique) , Quantum Monte Carlo and cluster
methods.  In this study, we use Cluster Dynamical Mean Field Theory (CDMFT) \cite{kotliar2} to compute the one-particle Green function.   In this method, we consider exactly the dynamics on a 4-site (plaquette) cluster and treat the rest of the lattice as a bath whose degrees of freedom are integrated over. The coupling of the cluster to the bath is thus treated in a mean-field fashion and the cluster quantities are evaluated self consistently using the cumulant lattice reconstruction scheme\cite{stanescukot}. While no variational principle exists for this method, several tests in limiting cases in which the answers are known exactly have revealed that the local correlations that the CDMFT method builds in provide an adequate starting point for quantifying the physics of strong correlations.  

Using the CDMFT method, we obtained the one-particle spectral function $A({\bf k} , \omega)$ for various values of hole doping ${\bf x}$ and on-site Coulomb repulsion ${\bf U}$ (expressed in units of the nearest neighbor hopping integral ${\bf t}$) . Hole dopings in the range of 0.03 (lightly doped Mott insulator) up to around 0.30 (heavily overdoped superconductor) have been considered together with a maximum ${\bf U}$ value of 12$t$.
 \begin{figure}
 
 \includegraphics[width=8cm]{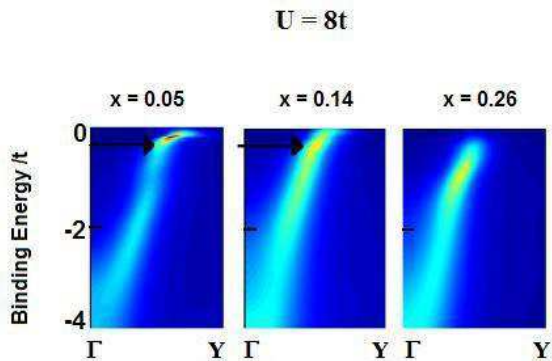}
 \caption{Intensity plot of the spectral function computed along the nodal $\Gamma=(0,0) \rightarrow Y=(\pi , \pi)$ direction for U = 8t at three different hole dopings. The  arrows in the first two panels indicate the position of the kink.}
\label{fig1}
 \end{figure}
 Typical spectral functions, $A(\textbf{k},\omega)$, obtained 
from CDMFT are displayed in three panels in Fig. ~\ref{fig1}.  We have plotted here only the spectral function intensity for different hole dopings in which the brightness at each point $({\bf k} ,\omega)$ is indicative of the magnitude of $A({\bf k} ,\omega)$ at that point. The bright band in each panel is the region of maximum $A({\bf k} ,\omega)$ and defines the $\omega$ vs ${\bf k}$ dispersion curve. The most interesting feature in these plots is the presence of a distinct kink in the dispersion curve at low doping values and an absence of the kink at the highest doping.  The overall doping dependence is in rough agreement with the experimental trends. Also shown in Fig. ~\ref{fig2} is the U-dependence of the kink energy for three  $U=4\bf{t}$ , $U=8\bf{t}$ and $U=12\bf{t}$. As seen in Fig.~\ref{fig2}, the kink energy scales inversely with ${U}$. Fitting the kink energies to $1/{U}$ gives a very good linear plot with as slope approximately 4. In other words the kink energy scale is given by $4 {t^2}/{U}$. 

\begin{figure}[h]
\includegraphics[width=8.2cm]{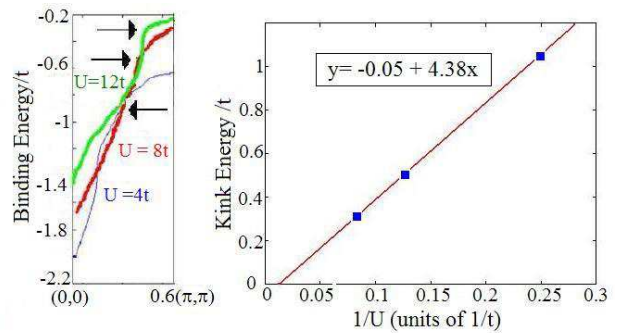}
 \caption{Left panel shows dispersion curves at $x=0.05$ plotted for 3 different $\bf{U}$ values with the black arrows showing the position of the kink in each curve. Right panel shows kink energy plotted vs 1/U and its linear fit. }
\label{fig2}
 \end{figure}

\begin{figure}

\includegraphics[width=7cm]{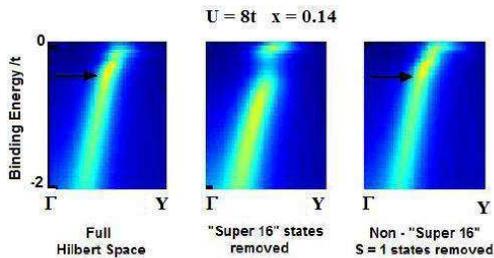}
\caption{Spectral function for $x=0.14$ obtained in two restricted Hilbert spaces on the plaquette (center and right panels) compared with the complete spectrum (left panel). The kink goes away only when the 'Super 16' states are removed from the Hilbert space and reappears when they are included even when other states with $S_z$ =$\pm$ 1 are absent. }

\label{fig3} 
\end{figure}

To explore the physics of the kink in further detail, we looked for
specific short-range electron correlations within the plaquette that
give rise to the kink.  Such details are readily available from the
176 local resolvents that comprise the impurity solver method we
employed, namely the non-crossing approximation (NCA). To achieve
this,  NCA has been selectively employed on various subspaces of the
full Hilbert space of the plaquette in order to isolate the effect of
each subspace on the spectral function. This requires an extensive
search on the $4^4$ = 256 dimensional Hilbert space and  isolating the
states relevant to kink formation. It turns out that out of 256
plaquette states, only 16 are involved in giving rise to the kink. All
of these states ( labelled `Super 16' in Fig. ~\ref{fig3} )  have 4
electrons with a total spin $S_z$ = $\pm$1 i.e. there are three
same-spin electrons and an opposite spin electron on a plaquette.  All
of the states in the `Super 16' subspace have some doubly occupied character. By doubly occupied character, we simply mean that there is a non-zero overlap with the doubly occupied sector not that the wavefunction is predominantly doubly occupied. We also
verified that the kink vanishes if the doubly occupied sector is
eliminated.  Consequently, the kink we have found here does not have a
simple interpretation in terms of spin fluctuations as such a scenario
would not require the explicitly doubly occupied sector. A kink
arising from an explicit mixing between singly and doubly occupied
sectors with $S=1$ is consistent, however, with the physics mediated
by the charge 2e boson\cite{ftm1,ftm2,ftm3} that appears in the exact
low-energy theory of a doped Mott insulator modeled by the Hubbard
model.  Excitation of the boson, which mediates mixing with all the
doubly occupied sectors, occurs on the energy scale $t^2/U$. Once the
boson is excited, the electron dispersion should change. Hence, we conclude that our numerical results are consistent with the physics that the charge 2e-boson\cite{ftm1,ftm3} mediates at low energy in the Hubbard model.
 
\begin{figure}[h]
 \includegraphics[width= 8.0cm]{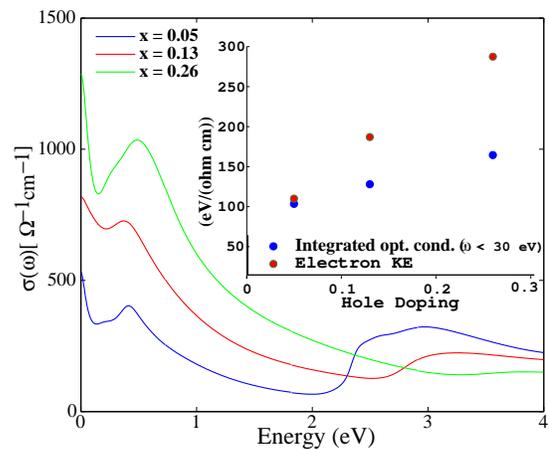}
 \caption{(Color online).Optical conductivity plots computed using CDMFT for three different hole dopings for U = 8t, T = 0.1t. Inset: Integrated optical conductivity $K(\Omega)=\label{sum}
 \int_{0}^{\Omega} \sigma(\omega) d\omega$, up to energy of $(\Omega =60t=30eV)$ (blue dots) and electron kinetic energy (red dots)plotted for three different hole dopings for U = 8t, T = 0.1t.}
 \label{fig4}
 \end{figure}

\begin{figure}[h]
 \includegraphics[width=8.0cm]{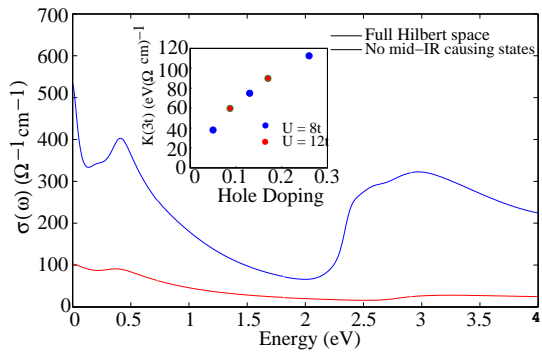}
 \caption{Optical conductivity plot for $U = 8t$, n = 0.94, T = 0.1t
   with distinct  mid-IR peak contrasted with the contribution from a
   restricted Hilbert space with certain S=0 states removed. Inset:
  Integrated optical conductivity $K(\Omega)=
 \int_{0}^{\Omega} \sigma(\omega) d\omega$, up to $(\Omega=3t=1.5eV)$ (restricted f-sum
  rule) for $U=12t$ and $U=8t$.  }
 \label{fig5}
 \end{figure}

\begin{figure}
\centering
\includegraphics[width=4.2cm]{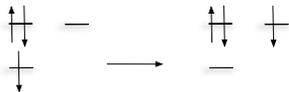}
\caption{Hopping process that mixes the upper and lower Hubbard bands
  that leads to the mid-IR band.  This process requires that the hole and the doubly occupied site are neighbours.  It is the motion of this bound state that we attribute to the mid-IR band.}
\label{midir}
\end{figure}

We also computed the optical conductivity.  To obtain a direct link between the conductivity and the spectral function, we worked in the non-crossing approximation 
\beq
\sigma(\omega) &=& \frac{2 e^2 }{\nu \hbar^2 N} \displaystyle
\sum_{\bf k} (\nabla_{\bf k} \epsilon_{\bf k})^2 \int \frac{d
  \omega}{2 \pi} A(\bf{k},\omega ^\prime)\nonumber\\
&&\times A(\bf{k},\omega ^\prime + \omega) \frac{f(\omega ^\prime) - f(\omega ^\prime + \omega)}{\omega}
\label{cond}
\eeq 
to the Kubo formula for the
conductivity \cite{jarrell, kotliar3,prelovsek}where $f(\omega)$ is the Fermi distribution function and $A(\omega,k)$ is the spectral function. 
 Here  $\nu$ is the unit cell volume and $\epsilon_{\bf k}$ is the dispersion for the non - interacting system. The optical conductivity has been computed for various dopings and $\bf U$ values. Fig. ~\ref{fig4} displays the optical conductivity for three different
hole dopings. The optical conductivity plots show a  peak-like
resonance feature  (Fig. ~\ref{fig4}) at 0.5 eV
(4000 $cm^{-1}$) for  $U = 8t$ which  falls right inside the mid- IR
region where the
experimental\cite{tanner,cooper,uchida,opt1,opt2,opt3} data shows an
absorption peak. The magnitude of the optical conductivity is also in perfect agreement with experiment\cite{cooper,uchida}. The physical origin of this peak can be determined by
focusing (see Fig. (\ref{fig5})) on the resolvents for states in the plaquette that contribute
significantly to the optical conductivity.  Surprisingly, of the 256 plaquette states, only a single
state 
in the N=4 and $S_z=0$ sector contributes (see Fig. (\ref{fig5})) to the mid-IR peak.  This state
has three key characteristics: 1) it contains a mixture of singly
 (87$\%$) and doubly occupied (13$\%$) sites, 2) its
energy is $-1.3t$, essentially the energy of mid-IR peak, and 3) the
spatial symmetry of the eigenstate is d$_{x^2-y^2}$.  Any physical
process which meets these constraints must also couple to the charge
not simply to the spin sector as in the case of
antiferromagnetic spin fluctuations.  A further hint as to the origin
of the MIB is the calculation of Haule and Kotliar\cite{haulek} indicating that
the MIB is absent in the traditional implementation of the $t-J$ model
unless superconductivity is present.  By traditional implementation, we mean the $U=\infty$ limit
in which $J$ is (inconsistently) treated as being finite but double occupancy is
excluded in the Hubbard basis rather than being excluded only from the transformed
basis\cite{eskes}.  The typical process, shown in Fig. (\ref{midir}), which is eliminated by
imposing the artificial $U=\infty$ constraint
results in the mixing of the high and low energy scale by
virtue of a hole neighbouring a doubly occupied site. This resonance
persists even at $x=0$ (though with perhaps vanishing weight) and hence we predict that the restricted f-sum
rule at finite doping should extrapolate to a non-zero value at $x=0$
as long as $U\ne\infty$ as depicted in the
inset of Fig. (\ref{fig5}).  As this is the only process through
$O(t^2/U)$ that is left out in the $U=\infty$ limit, we conclude that
the resonance shown in Fig. (\ref{midir}) is the origin of the
mid-IR. This would indicate that the mid-IR is strongly momentum
dependent (and hence not adequately described by single-site
analyses\cite{millis}), having $d_{x^2-y^2}$ symmetry, our principal conclusion. That the process in Fig. (\ref{midir}) causes the mid-IR was also the result of our analysis\cite{ftm3} of the exact low-energy theory of the Hubbard model.  

An explicit assumption in the  NCA procedure (Eq. ~\ref{cond}) is
that vertex corrections are not important. That is, the current-current correlator is approximated by the simple bubble diagram and higher order terms (vertex corrections) are neglected.  This approximation is only correct in the limit of infinite dimensions\cite{khurana}.  To determine how valid this approximation is for a d=2 system, we computed the kinetic energy explicitly versus the value predicted from the sum rule
 \beq K(\infty) = - \frac {\pi e^2 a^2}{2 \hbar^2 V} E_{\it{Kin}}
 \eeq
 where $K(\Omega)=
 \int_{0}^{\Omega} \sigma(\omega) d\omega$ is the integrated optical spectral weight, \textit{e} is the electron charge, \textit{a} is the lattice
 constant and \textit{V} is the volume of the unit cell. This sum rule
 is expected to hold as long as the hopping terms are mediated by
 nearest-neighbour hopping processes.  The integrated optical
 conductivity and the electron kinetic energy (computed explicitly
 using CDMFT) have been plotted for various hole doping levels in the
 inset of Fig.~\ref{fig4}. We find that the sum rule is  reasonably
 well obeyed for low hole dopings (x = 0.05) near
 half-filling. However, further away from half-filling where the
 low-energy sector contains increased spectral weight transferred from
 high energy, the integrated optical conductivity fails to account for
 the kinetic energy.  This behaviour is not unexpected as the vertex corrections ignored in Eq. (\ref{cond}) are strongly doping dependent and hence Eq. (\ref{cond}) is more accurate close to half-filling. As our focus is on the the mid-infrared feature which is most well-defined close to half-filling, our treatment here should be adequate.

The key feature that distinguishes this work is the finding that double occupancy mediates both the low-energy kink and the mid-IR band.   Our result that the mid-IR is caused by a
single state (of the 256 plaquette states) that has $d_{x^2-y^2}$
symmetry is striking and indicates that the mid-IR is
 likely to be caused by the same physics that
underlies the pseudogap as optics experiments seem to indicate\cite{basov2}. Further, the mechanism identified here, which is identical to that mediated by the the charge 2e boson in the exact low-energy theory of the Hubbard model\cite{ftm1,ftm2,ftm3} (Fig. (\ref{midir})) could explain why
an extrapolated non-zero intercept of the restricted f-sum rule might
be a generic feature of doped Mott insulators, in contrast to recent claims\cite{millis}.

\acknowledgements We thank A. Millis for sharing with us his DMFT results and for a discussion on the sum rule, G. A. Sawatzky, G. Kotliar and D. Basov for helpful discussions and NSF-DMR0605769 and NSF-PHY05-051164 for partially funding this work.


\begin{thebibliography}{1}
 \bibitem{lanzara} A. Lanzara \textit{et al.}, Nature \textbf{412} ,510 (2001).
\bibitem{zhou} X.J. Zhou \textit{et al.}, Phys. Rev. Lett. \textbf{95} , 117001 (2005).
 
 \bibitem{damascelli} A. Damascelli, Z. Hussain and Z-X. Shen, Rev. Mod. Phys. \textbf{75}, 473 (2003).
 \bibitem{bogdanov} P. V. Bogdanov \textit{et al.}, Phys.Rev. Lett. \textbf{85}. 2581 (2000).
 \bibitem{cuk} T.Cuk \textit{et al.}, Phys.Rev. Lett. \textbf{93}. 117003 (2004).
 \bibitem{johnson} P. D. Johnson \textit{et al.}, Phys.Rev. Lett. \textbf{87}. 117007 (2001).
  \bibitem{kaminski} A. Kaminski \textit{et al.}, Phys.Rev. Lett. \textbf{86}. 1070 (2001).
  \bibitem{kim} T. K. Kim  \textit{et al.}, Phys.Rev. Lett. \textbf{91}. 167002 (2003).
 
 \bibitem{valla} T. Valla , cond-mat/0507653.

\bibitem{vollhardt}K. Byczuk, \textit{et al.} Nat. Phys. {\bf 3}, 168 (2007).

\bibitem{graf}J. Graf, et al., Phys. Rev. Lett. {\bf 98}, 67004 (2007).

 \bibitem{meevasana} W. Meevasana \textit{et al.},Phys. Rev. B \textbf{75}, 174506 (2007).
 \bibitem{basov1} D. N. Basov and T. Timusk, Rev. Mod. Phys. \textbf{77}, 721 (2005).
 \bibitem{basov2} Y. S. Lee \textit{et al.} Phys. Rev. B \textbf{72}, 054529 (2005).
 \bibitem{meinders} M. B. J. Meinders, H. Eskes and G. A. Sawatzky, Phys. Rev. B    \textbf{48},3916-3926 (1993).
 
 \bibitem{tanner} S. W. Moore \textit{et al.}, Phys. Rev. B \textbf{66}, 060509 (2002).
 \bibitem{cooper} S. L. Cooper, \textit{et al.},Phys. Rev. B \textbf{41}, 11605 (1990).
\bibitem{uchida}S. Uchida \textit{et al.}, Phys. Rev. B {\bf 43}, 7942 (1991).
\bibitem{opt0}D. van der Marel \textit{et al.}, Nature {\bf 425}, 271 (2003).
\bibitem{opt1}S. W. Moore \textit{et al.}, Phys.Rev. B {\bf 46}, 605091 (2002).
\bibitem{opt2}J. Bouvier \textit{et al.}, Phys. Rev. B {\bf 45}, 8065 (1992).
\bibitem{opt3}Y. S. Lee \textit{et al.},Phys. Rev. {\bf 72}, 54529 (2005). 
 \bibitem{dagotto} E. Dagotto, Rev. Mod. Phys. \textbf{66}, 763 (1994).
 \bibitem{lorenzana1} J. Lorenzana and G.A. Sawatzky, Phys. Rev. Lett. \textbf{74}, 1867
(1995).
\bibitem{lorenzana2} J. Lorenzana and G. A. Sawatzky, Phys. Rev. B \textbf{52},
9576 (1995).
\bibitem{leggett} A. J. Leggett, Proc. Natl. Acad. Sci. USA \textbf{96}, 8365 (1999).
\bibitem{turlakov} M. Turlakov and Anthony J. Leggett, Phys. Rev. B \textbf{67},
094517 (2003).
\bibitem{emery1} V. J. Emery and S. A. Kivelson, Physica C \textbf{209}, 597 (1993).
\bibitem{emery2} V.J. Emery and S.A. Kivelson, Phys. Rev. Lett. \textbf{74} , 3253
(1995).
 \bibitem{kotliar} A. Georges, G. Kotliar, W. Krauth and  M. J. Rozenberg , Rev. Mod. Phys. \textbf{68}, 13 (1996).
 \bibitem{kotliar2} G. Kotliar, S. Y. Savrasov, G. Pálsson and G. Biroli ,Phys. Rev. Lett. \textbf{87}, 186401 (2001).
\bibitem{stanescukot}T. D. Stanescu and G. Kotliar, Phys. Rev. B {\bf 74}, 125110 (2006).
\bibitem{ftm1}R. G. Leigh, P. Phillips, and T.-P. Choy, Phys. Rev. Lett. {\bf 99}, 14604 (2007).
\bibitem{ftm2}T.-P. Choy, R. G. Leigh, P. Phillips, and P. D. Powell,
  Phys. Rev. B {\bf 77}, 014512 (2008).
\bibitem{ftm3}T.-P. Choy, R. G. Leigh, and P. Phillips, arxiv:0712.2841.
 \bibitem{jarrell} M. Jarrell, J. K. Freericks and T. Pruschke,  Phys. Rev. B \textbf{51}, 11704 (1995).
 \bibitem{kotliar3} M. J. Rozenberg, G. Kotliar,  H. Kajueter, G. A. Thomas , D. H. Rapkine, J. M. Honig and P. Metcalf, Phys. Rev. Lett. \textbf{75}, 105 (1995).
\bibitem{prelovsek}M. M. Zemljic and P. Prelovsek, Phys. Rev. B {\bf 72}, 75108 (2005).
\bibitem{haulek}K. Haule and G. Kotliar, Europhys. Lett. {\bf 77},
 27007 (2007); ibid, arxiv:0709.0019.
\bibitem{eskes}H. Eskes \textit{et al.}, {\bf 50}, 17980 (1994).
\bibitem{millis}A. Comanac, L. de Medici, M. Capone, and A. J. Millis, Nat. Phys. {\bf 4}, 287 (2008).
\bibitem{khurana} A. Khurana, Phys. Rev. Lett. \textbf{64}, 1990 (1990). 
  \end{thebibliography}
 \end{document}